\documentclass{article}

\usepackage{arxiv}

\usepackage[utf8]{inputenc} 
\usepackage[T1]{fontenc}    
\usepackage{hyperref}       
\usepackage{url}            
\usepackage{booktabs}       
\usepackage{amsfonts}       
\usepackage{microtype}      
\usepackage{lipsum}
\usepackage{graphicx}
\graphicspath{ {./images/} }

\title{A brief introduction to satellite communications for Non-Terrestrial Networks (NTN)}

\author{
  Jos\'{e} Alberto Hern\'{a}ndez\\
  Universidad Carlos III de Madrid\\
  28911 Madrid, Spain \\
  \texttt{jahgutie@it.uc3m.es} \\
  \And  
  Pedro Reviriego \\
  Universidad Polit\'{e}cnica de Madrid\\
  28040 Madrid, Spain \\
  \texttt{pedro.reviriego@upm.es} \\
}

\begin{document}

\maketitle

\begin{abstract}
    At present (year 2023), approximately 2,500 satellites are currently orbiting the Earth. This number is expected to reach 50,000 satellites (that is, 20 times growth) for the next 10 years, thanks to the recent advances concerning launching satellites at low cost and with high probability of success. In this sense, it is expected that next years the world will witness a massive increase in mobile connectivity thanks to the combination of 5G deployments and satellites, building the so-called Space-Terrestrial Integrated Network (STIN), thanks to the emergence of Non-Terrestrial Networks (NTNs). This document overviews the foundations of satellite communications as a short tutorial for those interested in research and development on Space-Terrestrial Integrated Networks (STIN) and Non-Terrestrial Networks (NTN) for supporting 5G in remote areas. 
\end{abstract}



\section{Introduction}
\label{sec:introduction}

In the USA, about 20\% of the population lives in rural areas, this accounts for about 97\% of the total land. This number grows to 28\% in Europe, and about 40\% world-wide. In many cases, fiber deployment does not reach rural areas (at least the last mile), since this results very expensive for network operators, hard to justify in terms of Average Revenue per User (ARPU). Indeed, it is estimated that every single meter of fiber connectivity costs approximately 100 US dollars. The largest share of this cost includes digging, trenching and the civil works in general~\cite{schneir_2014}. Satellites can be a good solution to provide broadband connectivity in those areas where fiber cannot reach (deep rural, seaside, desert, mountains, etc).


Indeed, in the past years, the research community has witnessed a race toward deploying different satellite constellations to provide connectivity to both rural and remote areas. This is mainly due to the cost reduction in launching the satellites themselves, approximately a few thousand USD per kg of mass~\cite{techno_economic_sats} for SpaceX Falcon 9. In this sense, it is estimated that approximately 2,500 satellites are currently orbiting the Earth, a number that is foreseen to grow to 50,000 within ten years~\cite{mckensey_2020}.

Essentially, while GEO and MEO constellations suffer from high two-way delays, in the order of several hundreds of milliseconds, with subsequent performance degradation of TCP protocols, LEO constellations can further reduce such delays to few tens of milliseconds. Furthermore, High-Altitude Platforms (HAPs) operating at 20~Km distance can even reduce RTTs to few milliseconds.

It is worth noticing that light travels at approximately 300,000~km/s through the air, while it does at 200,000~km/s over silica fiber. That is, the air is 50\% faster than silica fibers in terms of propagation delay. This translates into a propagation delay of $3.33~\mu s/km$ for free-space communications and $5~\mu s/km$ for fiber transmission. In fact, some authors claim that satellite communications can be faster than fiber in wide area scenarios above 1,000~km~\cite{Handley_2018,Handley_2019}, especially in those regions with difficult conditions for fiber deployment (i.e. desert, mountains, etc).

In particular, a number of companies have focused on deploying Low Earth Orbit (LEO) satellite constellations (between 500 - 1200~km altitude) since latency in these cases are moderate (few tens of milliseconds). In addition to providing coverage to rural areas, satellites can provide connectivity worldwide and are very resilient to natural disasters and wars. LEO satellite constellations can provide sufficient connectivity performance for Machine-Type Communications (MTC) and Mobile Broadband (MBB) in such remote areas~\cite{ngmn_ntn}, paving the way for Non-Terrestrial Networks (NTN) that complement existing Terrestrial Networks, both fixed and mobile. The authors of~\cite{guidotti_LTE,guidotti_5g} provide a summary of architectures and challenges to integrate LEO constellations in the 5G ecosystem and even 6G~\cite{aranity_2021}. A detailed survey on this matter is exhaustively studied in~\cite{ieeeaccess_survey}.

Four major companies are already deploying LEO satellite mega-constellations, namely Telesat, Tesla's Starlink, OneWeb and Amazon Kuiper. The authors in~\cite{DELPORTILLO2019123} provide a thorough comparison of the LEO constellations and features provided by these four major players, showing tens of milliseconds latency and average throughput in the order of Gb/s per satellite. Vertical applications like rural broadband, IoT applications like smart agriculture and animal tracking, environmental protection and public safety can represent interesting market opportunities to trigger further satellite developments.

This article provides a brief review on the basic principles and design requirements of satellite communications. This review contains several numerical examples to give the reader a better feeling of the uses and applications of satcoms. To this end, Section~\ref{sec:soa} briefly reviews the most important design aspects of Satellite Communications. Section~\ref{sec:stin} introduces current mega-constellations and existing projects for NTNs as of year 2023. 
Finally, Section~\ref{sec:conclusions} concludes this work with a summary of its main contributions.


\section{An overview of satellite communications}
\label{sec:soa}

\subsection{Orbits and propagation delay}

\begin{figure}[htbp]
\centering
\includegraphics[width=0.6\textwidth]{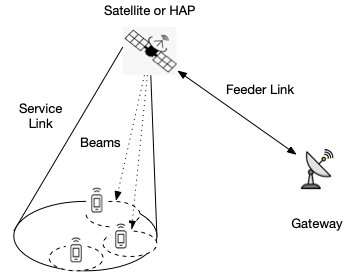}
\caption{Architecture and terminology.} 
\label{fig:arch}
\end{figure}

In general Non-Terrestrial Networks (NTNs) refer to networks providing connectivity through space-borne vehicles or airborne platforms, including satellites, High-Altitude Platforms (HAPs) and Low-Altitude Platforms (LAPs). These provide radio connectivity between the User Equipment (UE) on the ground and the vehicle which, in addition, provide connectivity to Terrestrial Networks (TN) through Ground Based Gateways (see Fig.~\ref{fig:arch}).

Depending on the altitude of the space-borne, multiple NTN options are possible:
\begin{itemize}
    \item Stationary satellites placed in GEO, operating at 35,876~km altitude. GEO scenarios are often equipped with Very High Throughput Satellites (VHTS), providing tens or hundreds of Gb/s capacity each. In this case, Doppler effects are negligible but propagation delays can reach up to several hundreds of millisecs for transparent satellites. 
    \item Non-stationary satellites positioned in MEO (7,000-25,000~km) or LEO (300-2,000~km), in relative motion to the earth. In these cases, latency values can be moderate, in the range of tens of milliseconds, but Doppler needs compensation. Satellite coverage and their cells may be stationary or not. The former requires the beams fixed on Earth, while the latter simply implies that the beams move at the same speed of the satellites (typically few km/s for LEO sats). In the case of non-stationary cells, methods for handover operations and roaming are required.
    \item High Altitude Platforms (HAPs) like planes or balloons, operating like satellites but closer to the Earth, at about 20~km distance. HAPs latency values are often below 10~ms.
    \item Low Altitude Platforms (LAPs) like drones or balloons at less than 1~km altitude.
\end{itemize}

The altitude of the space-borne has a clear impact on the round-trip time. In this regard, it is worth remarking that latency heavily affects TCP throughput in TCP/IP based networks, which for traditional TCP implementations is given by the Mathis formula~\cite{mathis}, further validated in~\cite{padhye_tcp}:
\begin{equation}
    \label{eq:mathis}
    Throughput_{TCP} < \frac{MSS}{RTT} \frac{C}{\sqrt{p_{loss}}}
\end{equation}
where MSS is the Maximum Segment Size and RTT is the end-to-end Round-Trip Time; $C$ is a constant that can be estimated from measurements (a number between 1 and 1.5 typically) and $p_{loss}$ is the packet loss probability, due to any factor (packet corruption, collisions in shared media or buffer overflow due to congestion).

\fbox{
  \parbox{\linewidth}{
    \textbf{Numerical example no. 1: } As an example, consider a connection between two cities separated $200~ms$, MSS of 1500~Bytes and packet loss probability of $10^{-9}$ (typical fiber loss)~\cite{turner}. Substituting in eq.~\ref{eq:mathis}, the maximum rate is 1.9~Gb/s (assuming $C=1$). If latency is doubled (i.e. $400~ms$), then the maximum TCP throughput drops to 950~Mb/s. On the other hand, if the packet loss probability is $10^{-6}$, then the throughput drops to 30~Mb/s for RTT values of $400~ms$. Thus, TCP throughput is heavily affected from both high-latency and unreliable links, especially the latter.
  }
}

\subsection{Frequency bands}

\begin{figure}[htbp]
\centering
\includegraphics[width=0.9\textwidth]{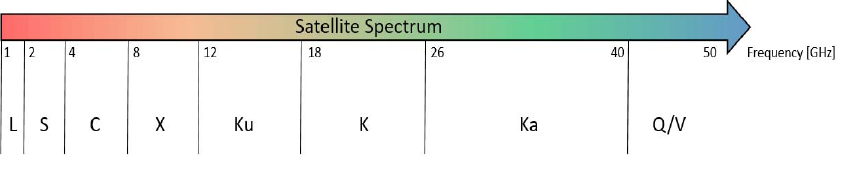}
\caption{Satellite spectrum~\cite{IEEE_surv_tutorial_kodheli}.} 
\label{fig:bands}
\end{figure}

Both Fig.~\ref{fig:bands} and Table~\ref{tab:freqbands} summarise detailed information regarding frequency bands allocated by the ITU for satellite communications. As shown, the L and S bands do not offer much bandwidth (tens or hundreds of KHz to few MHz typically) and are often destined to IoT applications with low bandwidth requirements. The Ka and Ku bands provide a lot more bandwidth (tens or hundreds MHz) and can be used to provide Mobile Broadband (MBB) connectivity, similar to those experienced in DSL connections (say 20~Mb/s) even similar to fiber to the home (100~Mb/s and above), especially in cases with high antenna gains. Finally, the Q/V bands offer even larger bandwidth capacity values (hundreds MHz to few GHz) but are more subject to atmospheric losses and absorption from rain. In this regard, some satellite companies are considering the V band for inter-satellite links (ISL) since they are above the clouds, offering mesh connectivity between satellites. Also, some experimental scenarios consider the W Band (between 75-110~GHz) which provides even more bandwidth than the Q and V bands, and should be also used for inter-satellite links (ISL) since this band heavily suffers from propagation impairments and rain fade.

\begin{table}[htbp]
    \centering
    \caption{ITU Frequency allocations for satellite communications}
    \begin{tabular}{|c | c | c |}
    \hline
    Sat Band & Downlink (DL) & Uplink (UL) \\
    \hline
    L band & 1518 – 1559 MHz & 1626.5 – 1660.5 MHz \\
    (GEO) & & 1668 – 1675 MHz \\
    \hline
    L band & & \\
    (Non-GEO) & 1613.8 – 1626.5 MHz & 1610.0 – 1626.5 MHz \\
    \hline
    C band & 3400 - 4200 MHz & 5725 - 7025 MHz \\
     & 4500 - 4800 MHz & \\
    \hline
    S Band & 2160 -2200 MHz & 1980 - 2025 MHz \\
    & 2483.5 - 2500 MHz & \\
    \hline
    Ku band & 10.7 - 12.75 GHz & 12.75 - 13.25 GHz \\
    & & 13.75 - 14.5 GHz\\
    \hline
    Ka band (GEO) & 17.3 – 20.2 GHz &  27.0 – 30.0 GHz\\
    \hline
    Ka band & 17.7 – 20.2 GHz & 27.0 – 29.1 GHz \\
     (Non-GEO) & & 29.5 – 30.0 GHz \\
    \hline
    Q/V band & 37.5 – 42.5 GHz   & 42.5 – 43.5 GHz, \\
    & 47.5 - 47.9 GHz & 47.2 – 50.2 GHz \\
    & 48.2 - 48.54 GHz &  50.4 – 51.4 GHz \\
     & 49.44 - 50.2 GHz & \\
    \hline
    \end{tabular}
    \label{tab:freqbands}
\end{table}

\subsection{Link-budget calculations}


Classical link budget calculations for satellite communications follow the well-known Friis propagation model, where the power at the receiver antenna $P_r$ (also referred to as signal strength $S$) is:
\begin{equation}
    \label{eq:friis}
    P_r = P_t \frac{G_tG_r \lambda^2}{(4\pi)^2 d^2} = S
\end{equation}
where $P_{tx}$ is the transmission power of the transmitting antenna, $G_t$ and $G_r$ are the transmission and reception gain of the two antennas, and $\lambda$ and $d$ are the transmission wavelength and slant range between the transmitter and receiver in the satellite link. Often, the product $P_tG_t$ is called the EIRP or Effective Isotropic Radiated Power. 

The receiving antenna both collects the above signal power $S$ and noise $N$. The amount of noise collected follows:
\begin{equation}
    N = k_B T B_w
\end{equation}
where $k_B$ is the Boltzmann constant ($1.380649\times 10^{-23}~m^2\cdot kg\cdot s^{-2}\cdot K^{-1}$ or $-228.6~\frac{dBW/K}{Hz}$), $T$ is the noise temperature and $B_w$ is the bandwidth of the receiving filter. Often, receiver systems and amplifiers are provided with noise figure (NF) values, which is a classical figure of merit used for both receivers and amplifiers (typically 5 to 9~dB of noise figure values). The noise temperature $T$ can be computed from the noise figure as:
\begin{equation}
    T = T_{ref} (10^{\frac{NF}{10}}-1)    
\end{equation}
where the reference (ambient) temperature $T_{ref}$ is often assumed 290~K (i.e. 16.85~ºC).

With these values of signal strength $S$ and noise power $N$ at the receiver, and neglecting interference, the signal-to-noise ratio (SNR) in dB follows~\cite{kodheli_linkbudget}:
\begin{eqnarray}
\nonumber SNR & = & EIRP~(dBW) \\
\nonumber & + & G_r/T~(dBi/K) \\
\nonumber & - & FSPL~(dB) \\
\nonumber & - & AtmLoss~(dB) \\
\nonumber & - & AdLoss~(dB) \\
\nonumber & - & B_w~(dBHz) \\
& - & K_B~(\frac{dBW/K}{Hz})) 
    \label{eq:linkbudget}
\end{eqnarray}
where $G_r/T$ is the reception's antenna figure of merit, that takes into account both reception Gain (dBi) and Noise Temperature (Kelvin). As an example, the following list gives an overview of typical terminal equipments and their figures of merit:
\begin{itemize}
    \item 3GPP Class 3 UE, with 0~dBi antenea gain (linear polarized), 200~mW (i.e. 23~dBm) transmission power and 7 or 9~dB Noise Figure. Assuming $NF=7~dB$ and ambient temperature $T_{ref}=290~K$, the noise temperature is $T=1163.4~K$, and $G_r/T=0~dBi-10\log_{10}(1163~K)=-30~dB/K$ at the receiver.
    \item Very Small Aperture Terminal (VSAT) with 12~dBi antenna gain (circular), 2~W transmission power and 5~dB Noise Figure. In this case, at the receiver $G_r/T=12-10\log_{10}(627~K)=-16~dB/K$ at the receiver.
    \item IoT devices with 0~dBi antenna gain, 290~K noise temperature and transmission EIRP = 23~dBm. The resulting $G_r/T=-24.6~dB/K$ at the receiver.
\end{itemize}

Concerning FSPL, AtmLoss and AdLoss, these refer to Free-Space Path Loss, Atmospheric Loss (due to gases, rain fade, etc) and any other Additional Loss respectively. FSPL is computed as follows:
\begin{equation}
    FSPL = 10\log_{10} \left( \frac{4\pi d f}{c}\right)^2 = 20\log_{10} \left( \frac{4\pi d f}{c}\right)
\end{equation}
where $f$ is the transmission frequency (as shown in Table~\ref{tab:freqbands}) and $d$ is the slant range, given by:
\begin{equation}
    d = -R_E \sin(\alpha) + \sqrt{R_E^2\sin(\alpha)^2 + h_s +2R_Eh_s}
\end{equation}
where $R_E$ refers to the Earth radius (6,371~km), $h_s$ is the satellite height/altitude and $\alpha$ is the elevation angle. The slant range $d$ is the distance from the user device to the satellite and can be often approximated by the satellite's orbit, especially for MEO and GEO satellites:
$$d = \sqrt{h_s^2 + (h_s \tan(\alpha))^2}$$

The atmospheric and additional losses take into account the attenuation due to absorption of different molecules in the atmosphere, mainly oxygen and water. Rain fade and availability, not taken into account in eq.~\ref{eq:linkbudget} accounts for the attenuation due to traversing clouds, rain, etc, which may reduce the availability of the links below 99\%. In this sense, the Crane model is often used to estimate these attenuation values on different weather environments (Tundra, Taiga, Maritime, Continental, etc)~\cite{crane_model}. Typically, link budget calculations consider clean sky assumptions (i.e. null attenuation due to rain and fading), but a margin value between 2 and 10~dB is often recommended to compensate from rain fading and other unexpected sources of power loss and attenuation.


\fbox{
  \parbox{\linewidth}{
    \textbf{Numerical example no. 2:} Consider a link between a ground station and a MEO satellite operating at $h_s=21000~km$ with elevation angle $\alpha=1.2^o$ (or 0.021~rad). 

    In this setting, we can approximate the slant angle with the height of the satellite, that is,     $d\approx h_s = 21,000~km$ since:
    $$d = \sqrt{21,000^2 + (21,000 \tan(0.021))^2} = 21,003~km$$

    The power transmission of the satellite is 26.6~Watt (or 14.4~dBW) and uses a helix antenna with 13~dBi gain. Assuming that the receiving station is a 3GPP Class 3 User Terminal, then the reception gain is $0~dBi$. Finally, the satellite operates in the S band, that is, the center transmission frequency is $f=2~GHz$, that is $\lambda = \frac{c}{f} = 0.15~m$. 

    In linear units:
    \begin{eqnarray}
    \nonumber G_t & = &  10^{\frac{13}{10}} = 19.95 \\
    \nonumber G_r & = & 10^{\frac{0}{10}} = 1 
    \end{eqnarray}

    The received power at the terminal for satcoms are typically very small, specially for MEO and GEO satellites. In this case, such received power follows~\ref{eq:friis}:
    $$P_r = 26.6 \frac{19.95\cdot 1\cdot  (0.15^2)}{(4\pi)^2 (21000 \cdot 10^3)^2} = 1.71\cdot 10^{-16}~W$$

    This satellite link uses the S band for transmission (2~GHz center frequency) and uses 1~KHz of bandwidth (that is, 30~dBHz). 

    Concerning SNR, the different values in eq~\ref{eq:linkbudget} are:
    \begin{eqnarray}
    \nonumber EIRP & = &  P_{tx} G_T = 10\log(26.6)+13 = 27.4~dBW \\
    \nonumber G_r/T & = & 0~dBi - 10\log_{10}(290\cdot(10^{\frac{7}{10}-1})) = -30~dB/K \\
    \nonumber FSPL & = & 20\log_{10}\left( \frac{4\pi (21000\cdot 10^3)(2\cdot 10^9)}{3\cdot 10^8} \right)^2 = 184.9~dB  \\
    \nonumber AtmLoss + AdLoss & = & 9.6~dB \\
    \nonumber B_w & = & 10\log_{10}(10^3) = 30~dBHz \\
    \nonumber K_B & = & -228.6~\frac{dBW/K}{Hz}
    \end{eqnarray}
  }
}

\fbox{
  \parbox{\linewidth}{
    \textbf{Numerical example no. 2 cont:}
    Hence, the Signal-to-Noise ratio follows:
    $$SNR = 27.4+(-30)-184.9 - 9.6 - 30-(-228.6) = 1.5~dB$$
    or $snr = 10^{\frac{1.5}{10}} = 1.41$ in natural units; that is, the signal power is 41\% greater than the noise power at the receiver. 
    
    This translates into a maximum spectral efficiency (SE) or $\beta_{max}$ of:
    $$\beta_{max} = \log_2(1+snr) = 1.27~bps/Hz$$
    This is further explained in the next section.
  }
}

\subsection{Bitrates, Shannon's capacity limit and Adaptive Coding and modulation}

After a given SNR is obtained from the link-budget analysis following eq.~\ref{eq:linkbudget}, this value together with the bandwidth used for transmission provides an upper bound of the maximum achievable bit rate $R_{max}$, as it follows from the Shannon-Hartley's theorem:
\begin{equation}
    R_{eff} < R_{max} = B_w \cdot \log_2(1+snr) = B_w \cdot \beta_{max}
    \label{eq:shannon}
\end{equation}
where the effective bitrate $R_{eff}$ used in transmission cannot be larger than the Shannon's limit $R_{max}$. 

The value $\beta_{max} = \log_2(1+snr)$ (in bps/Hz) is often referred to as spectral efficiency (SE) and measures how much bitrate can be obtained from a given bandwidth. Also:
\begin{equation}
    \label{eq:shannon_beta}
    \beta_{eff} < \beta_{max} = \log_2(1+snr)
\end{equation}

Typical spectral efficiency values range between 0.5 and 2~bps/Hz, reaching even up to 4 bps/Hz in some specific scenarios. Above 5~bps/Hz is often very difficult to achieve in satcoms, unless they are very close to Earth. 

\fbox{
  \parbox{\linewidth}{
    \textbf{Numerical example no. 3:} In the previous satcom link (1~KHz of bandwidth), the maximum achievable capacity follows:
    $$R_{max} = B_w \log_2 (1+snr) = 1\cdot 10^3 \log_2(1+10^{\frac{1.41}{10}}) = 1.27~Kb/s$$
   }
}

Taking the Shannon's limit the other way around, the communications link must provide sufficient SNR above the minimum required for a given spectral efficiency:
\begin{equation}
    snr_{eff} > snr_{req} = 2^{\beta_{max}}-1
\end{equation}
It is often recommended that a designed SNR provides a margin of a few dB above the Shannon's limit $SNR_{req}$ as a rule of thumb, to account for unexpected situations with SNR drop (atmospheric conditions, etc). 

Ideally, in the case of absence of noise, the spectral efficiency $\beta$ would only depend on the modulation used, its coding and reception filter roll-off. However, in the presence of noise, each modulation and coding scheme provides a different spectral efficiency as long as a minimum SNR is guaranteed. Table~\ref{tab:acm} shows the SNR requirements to achieve Quasi-Error Free (QEF) for some classical modulation and coding schemes (MODCOD) used in satellite links~\cite{kymeta}. As shown, low-order modulations like APSK are less efficient in terms of bits/symbol than higher-order ones, but their SNR requirements are also smaller. Here, QEF refers to only 2 errors for every 10,000 transmitted bits after Viterbi decoding (i.e. $BER = 2\cdot 10^{-4}$).

\begin{table}[htbp]
    \centering
    \caption{MODCOD table for a theoretical DVB modem, based on Shannon's limit~\cite{kymeta}}
    \begin{tabular}{|c c c|}
    \hline
    MODCOD & SE (bps/Hz) & SNR for QEF (dB) \\
    \hline
    APSK 1/2 & 0.4 & -2 \\
    CPSK 1/4 & 0.5 & 0 \\
    CPSK 1/2 & 0.6 & 1 \\
    CPSK 3/4 & 0.65 & 2 \\
    DPSK 1/4 & 0.75 & 3 \\
    DPSK 1/2 & 0.9 & 4 \\
    DPSK 3/4 & 1.05 & 6 \\
    DPSK 5/6 & 1.25 & 7 \\
    DPSK 7/8 & 1.5 & 9 \\
    \hline
    \end{tabular}
    \label{tab:acm}
\end{table}

Typically, modems have a wide range of available modulation and coding (MODCOD) schemes that can be used depending on the SNR link budget, which can be dynamically adjusted depending on the conditions of the satellite link. 

\fbox{
  \parbox{\linewidth}{
    \textbf{Numerical example no. 4}: As an example, consider we want to design a MODCOD scheme for the previous satcom link, where the SNR obtained was $1.41~dB$. Looking at Table~\ref{tab:acm}, CPSK 1/2 requires a minimum SNR of 1~dB, and we have 0.41~dB as margin. This MODCOD scheme offers an spectral efficiency of $0.6~bps/Hz$ which is much smaller than the maximum theoretical value given by the Shannon-Hartley ($\beta_{max} = 1.27~bps/Hz$). Thus, the effective bitrate achieve with quasi-error free (QEF) performance is $0.6\cdot 1~Kb/s = 600~bps$.
  }
}

\subsection{Increasing capacity with multiple beams per satellite and frequency reuse}

At present, at least four major private companies (Amazon Kuiper, Oneweb, Telesat and Starlink) are in the process of deploying large LEO satellite constellations with hundreds (even thousands) satellites at few hundred km above Earth surface. Some of the features of these four mega-constellations are explained in ~\cite{hypatia} and summarised in Table~\ref{tab:constellations}.

\begin{table}[htbp]
    \centering
    \caption{Constellations for Starlink, Kuiper and Telesat~\cite{hypatia}}
    \begin{tabular}{|c c c c c c |}
    \hline
    & Shell & Height (km) & Orbits & Sats/orbit & inclination $\alpha$ \\
    \hline
    Starlink & S1 & 550 & 72 & 22 & 53$^o$ \\
     & S2 & 1,110 & 32 & 50 & 53.8$^o$ \\
     & S3 & 1,130 & 8 & 50 & 74$^o$ \\
     & S4 & 1,275 & 5 & 75 & 81$^o$ \\
     & S5 & 1,325 & 6 & 75 & 70$^o$ \\
    \hline 
    Kuiper & K1 & 630 & 34 & 34 & 51.9$^o$ \\
    & K2 & 610 & 36 & 36 & 42$^o$ \\
    & K3 & 590 & 28 & 28 & 33$^o$ \\
    \hline
    Telesat & T1 & 1,015 & 27 & 13 & 98.98$^o$ \\
    & T2 & 1,325 & 40 & 33 & 50.88$^o$ \\
    \hline
    \end{tabular}
    \label{tab:constellations}
\end{table}

The footprint area $A_{sat}$ covered by one single satellite follows:
\begin{equation}
    D_{sat} = \frac{P_{Earth}}{N_{orbit}}
\end{equation}
where $P_{Earth}$ and $N_{orbit}$ are the Earth perimeter (40,075~km) and number of satellites per orbit respectively. 

\fbox{
  \parbox{\linewidth}{
    \textbf{Numerical example no. 5:} Consider Shell S1 of Starlink, with 22 satellites per orbit. This means that each satellite covers a diameter of:
    $$D_{sat} = \frac{40,075~km}{22} =1821.6~km\textrm{ of diameter}$$
    The area/footprint covered per satellite is then: 
    $$A_{sat}= \pi \left( \frac{D_{sat}}{2}\right)^2 = 2.6\cdot 10^6~km^2$$ 

    Since the total Earth surface is $Su_{Earth}=510.1\cdot 10^6~km^2$, then the 22 sats cover only 11\% of the total Earth surface. Shells S2-S5 should cover the rest of the Earth. 
  }
}


Each LEO satellite is often equipped with multiple antenna beams pointing at different regions (or cells) in its footprint area, and allowing frequency reuse like in mobile networks. This can be achieved in multiple ways, a typical one is by using phased array antennas with high directivity. Indeed, High or Very-High Throughput Satellites (HTS/VHTS) represent an evolution of satellites towards higher capacity through more spot beams and higher frequency reuse. These, applied in LEO orbit constellations, can further provide high bandwidth and reduced latency to enable Mobile Broadband (MBB) and Machine-Type Communications (MTC) in places where both fibre and 5G connectivity has limitations (deep rural areas, sea-side, mountains, etc). V/HTS can be identified by two key technological features~\cite{guan_review}:
\begin{itemize}
    \item The use of multiple spot beams (tens, even hundreds) of narrow beams covering a small geographical area cells, as shown in Fig.~\ref{fig:arch}. 
    \item The frequency reuse of allocated bandwidth in non-adjacent beams/cells, thus higher throughput of the satellite.
\end{itemize}

Indeed, more capacity can be provided to a given region by partitioning it into smaller sub-regions or cells covered by individual spot beams and leveraging frequency reuse. In this sense, capacity can scale up in the same way as in mobile networks by re-using multiple times the same frequency on non-adjacent cells, while keeping the the Signal to Noise and Interference Ratio (SINR) under acceptable limits for digital communications. Thus, the total satellite capacity $R_{tot}$ increases with the number of beams and polarizations as: 
\begin{equation}
    R_{tot} = \beta B_w \left( \frac{N_p N_b}{N_c} \right)  \cdot (1-\eta_{guard})
\end{equation}
where $N_p$ stands for the number of polarizations (1 or 2), $N_b$ is the number of spot beams (several tens, even hundreds for VHTS), $N_c$ is the number of colors or frequencies (3, 4 or 6 typically), and $\eta_{guard}$ is the guard-band between sub-bands (often a value between $5-10\%$).

\fbox{
  \parbox{\linewidth}{
    \textbf{Numerical example no. 6:}  Consider a satellite operating in the Ku-band with 1.5~GHz bandwidth and spectral efficiency of 2~bps/Hz. Under the assumption of 2 polarizations, 7 colors and 60 spot beams, the total capacity delivered by this satellite in the service link is up to:
    $$R_{tot} = 2 \cdot 1.5\cdot 10^9 \frac{2\cdot 60}{7} \cdot 0.9 = 46~Gb/s$$
    }
}

Indeed, in the Ku and Ka bands, bandwidth values per spot beam of 1.5-2 GHz are possible. The largest Ka-band satellites are Jupiter-2 and ViaSat-2, offering total aggregate capacity values of 200 to 300 Gb/s.

However, it is worth remarking that using multiple spot beams on-board heavily increases the size and weight of the satellite. For instance, a one hundred spot beams may account for 2,000~Kg of mass~\cite{future_ku}. As a rule of thumb, one Kg of weight can cost around 1,000 USD to get it in the sky\footnote{"Launch costs to low Earth orbit, 1980-2100", \url{https://www.futuretimeline.net/data-trends/6.htm}, last access February 2023}. In this light, scaling HTS satellites to VHTS is cost effective since the cost per Gb/s decreases following a power-law in these types of satellites, empirically~\cite{guan_review}:
\begin{equation}
Cost = 167.3\cdot (R_{tot})^{-0.886}
\end{equation}

In general, when using multiple beams, each single beam may interfere with adjacent ones in the frequency of operation. The final Signal to Interference and Noise Ratio (SINR) is obtained from combining both noise and interference as:
\begin{equation}
    SINR = \frac{S}{N+I} = \frac{1}{\frac{1}{SNR}+\frac{1}{SIR}}
\end{equation}
Thus, the network designer must be careful at balancing both noise and interference to not reach important signal degradation.

Essentially, the antennas are often designed with high directivity to well illuminate a given cell, while the sidelobes that may appear in neighbouring cells are well below the main lobe, typically 10~dB lower or above. In this sense, a given cell may receive power from other interfering cells, but such interfering power should be very low. 

\fbox{
  \parbox{\linewidth}{
  \textbf{Numerical example no. 7:}  For instance, consider a LEO sat illuminating a cell with SNR = 9~dB (i.e. the signal is 8 times stronger than the noise power). In this cell, the SIR observed from other interferring cells is only 6~dB (i.e. the signal power is 4 times stronger than the interfering adjacent signals). Then, the combined SINR reduces to 2.67 times or 4.25~dB:
  $$SINR = \frac{1}{\frac{1}{9}+\frac{1}{8}}= 2.67\textrm{ or }4.25~dB$$
  However, if the SIR is 12~dB, then the new SINR becomes 5.14~dB (from SNR = 9~dB).
  }
}

\subsection{Phased Array antennas, radiation pattern and directivity}

To achieve highly directive antennas on board, the use of linear or planar phased arrays of $N$ elements is often considered~\cite{array_article}. As an example, consider the Equally-Space Linear Array (ESLA), where $N$ array elements are separated by some distance $d$, for a total distance of $D=(N-1)d$. In this case, the Array Factor (AF) follows: 
\begin{equation}
    AF_{ESLA}(\psi) = e^{j(N-1)\frac{\psi}{2}} \frac{\sin\left(N\frac{\psi}{2}\right)}{\sin\left(\frac{\psi}{2}\right)}
\end{equation}
where $\psi = kd \cos(\theta)$; here the array is considered to receive signal from a plan wave incident at angle $\theta$ to the plane of the array. It is worth remarking that an isotropic antenna has $AF = 1$.

The maximum value of AF occurs when $\psi = 0$, resulting in $AF = N$, which is the directivity of this type of antenna:
\begin{equation}
    D_{ESLA} = N    
\end{equation}

\begin{figure}
    \centering
    \includegraphics[width=0.5\textwidth]{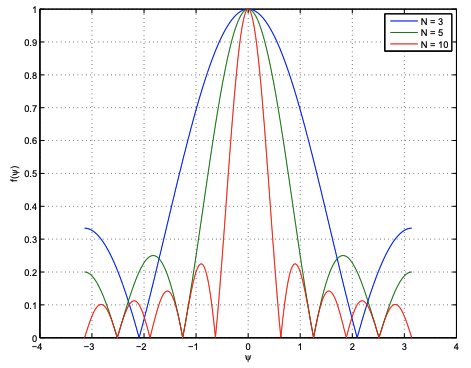}
    \caption{Plots of $|f(\psi)|$ for various $N$~\cite{array_article_toronto}.}
    \label{fig:AFs}
\end{figure}

Hence, disregarding the phase factor $e^{j(N-1)}$ and normalising, we obtain:
\begin{equation}
    f(\psi) = \frac{\sin\left(N\frac{\psi}{2}\right)}{N\sin\left(\frac{\psi}{2}\right)}
\end{equation}
which can be used to plot the radiation pattern and find the area of a cell covered by a beam constructed with an ESLA. Fig.~\ref{fig:AFs} shows examples of AF for different values of $N$.

\begin{figure}
    \centering
    \includegraphics[width=0.25\textwidth]{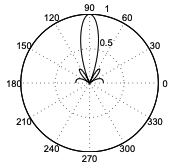}
    \caption{Radiation pattern for $N=5$ and $d=\frac{\lambda}{2}$~\cite{array_article_toronto}.}
    \label{fig:n5}
\end{figure}

As the number of array elements $N$ increases, the width of the main lobe in the radiation pattern decreases, making antennas more directive (with higher gain and smaller area covered). Also, as $N$ increases, the sidelobe level (SLL) decreases, producing less interferences on adjacent cells. For example, for $N=5$ and $d=\frac{\lambda}{2}$, the directivity of this array is $D_{ESLA}=N=5$ while the side lobes are very low (see Fig.~\ref{fig:n5}). A metric of interest is the Half-Power Beam Width (HPBW) that gives the angle at which the main lobe drops to one half (or 3~dB less directivity than the maximum).

Other arrays than ESLA have different properties and radiation patterns. For instance, Planar Rectangular Arrays (PRA) with $N$ elements again $\lambda/2$ spaced, have a maximum directivity of:
\begin{equation}
    D_{PRA} = N\pi    
\end{equation}

The gain of an antenna is typically smaller than its directivity by a factor $k_{ef}$ smaller than one:
\begin{equation}
    \label{eq:gain_directivity}
    G = k_{ef} D\quad \textrm{or}\quad G = D + 10\log_{10}(k_{ef})
\end{equation}
where efficiency $k_{ef}=\frac{P_{rad}}{P_{in}}$ accounts for the ratio of power effectively radiated to the air $P_{rad}$ divided by the input power to the antenna $P_{in}$. That is, not all incoming power into the antenna is transformed into radiation, part of it is lost; such loss is represented by $k_{ef}$, and is typically between 0.5 to 0.9. Remark that a value of $k_{ef}=0.5$ translates into a 3~dB difference between Gain and directivity, as noted in Eq.~\ref{eq:gain_directivity}.

It is worth remarking that Gain is relative to isotropic radiation, making the effective aperture of a given antenna proportional to its gain: 
\begin{equation}
    A_e = \frac{\lambda^2}{4\pi} G
\end{equation}
The ideal isotropic antenna radiates equally in all 3D directions, therefore it has no gain ($G=1$ in linear units or $G = 0~dBi$). 


In general for planar arrays, the directivity and the HPBW are related by the following approximation:
\begin{equation}
    D \approx \frac{32,400}{\theta_{HPBW\_1d}\theta_{HPBW\_2d}}
\end{equation}
where $\theta_{HPBW\_1d}$ and $\theta_{HPBW\_2d}$ are the angles (in degrees) where power drops 3~dB (or one half).


\begin{table}[htbp]
    \centering
    \caption{Directivity and HPBW for different Phased Array antenna configurations}
    \begin{tabular}{|c c c|}
    \hline
    Antenna & Directivity & HPBW \\
    \hline
    Isotropic & 1 (or 0~dBi) & -\\
    Linear $N=3$ & 3 (or 4.7~dBi) & 104º \\
    Linear $N=7$ & 7 (or 8.4~dBi) & 68º \\
    Linear $N=11$ & 11 (or 10.4~dBi) & 54º \\
    Planar $4\times4$ & $16\pi$ (17~dBi) & 25º \\
    Planar $8\times8$ & $64\pi$  (or 23~dBi)& 12.7º \\
    Planar $16\times16$ & $256\pi$ (29~dBi) & 6.4º \\
    Planar $32\times32$ & $1024\pi$ (35~dBi) & 3.2º \\
    \hline
    \end{tabular}
    \label{tab:array_examples}
\end{table}

Table~\ref{tab:array_examples} shows some examples of linear and planar arrays directivity and HPBW for different number of elements $N$. All cases assume untappered phased arrays, that is, uniformly weighted. As shown, narrow-beam antennas can also achieve high directivity and gain, but require phased arrays with multiple radiating elements. 

\fbox{
  \parbox{\linewidth}{
    \textbf{Numerical example no. 8: } Consider the case of a LEO satellite operating at 500~km altitude, willing to have on board multiple antenna beams, each beam covering an area of $7,854~km^2$ (that is, a circle with radius $R_{cell}=50~km$ or 100~km of diamater). Then, the HPBW of the antenna to illuminate that area should be:
    $$HPBW = 2\tan^{-1} \frac{R_{cell}}{h_s} = 2\tan^{-1} \frac{50}{500} =0.2~\textrm{rad or } 11.4^o$$
    Thus, looking at Table~\ref{tab:array_examples}, the designer decides to employ an 8x8 Planar Array antenna. Such an antenna has a directivity of 23~dBi on the center of the cell, and 3~dB less at the borders, i.e 20~dBi. Such antenna gain can be used in the Friis equation~\ref{eq:friis} to dimension the satellite link and further obtain the SNR and spectral efficiency of the link.
  }
}

Indeed, the directivity of the beams play an important role to properly cover its cell and not interfere adjacent ones where the same frequency is reused. Directivity increases with the number of antenna elements $N$, but also Side-Lobe Levels (SLL) reduce as $N$ grows, thus producing less interference in neighbouring cells. For instance, the SLL for a linear array with $N=3$ elements is 0.35 (i.e. -5~dB), while for $N=10$ is 0.22 (i.e. -6.6~dB). Typical SLL in modern phased arrays with high directivity often start on -10~dB onwards, thus limiting the interference contribution SIR to neighbouring cells.


\section{An overview of ongoing satellite constellations for Non-Terrestrial Networks}
\label{sec:stin}

In this section, we briefly overview some of the characteristics of ongoing NTN projects, including technical aspects like link budget, spectral efficiency, bandwidth and bitrate. More details can be found in~\cite{ngmn_ntn}.

\subsection{Thales Alenia Space: LEO constellation}

This project is intended to provide land-mobile connectivity to users (pedestrians walking, 3~km/h) in North America. The satellite antenna has 34~dBW/MHz of EIRP density and G/T of 1.1~dB/K while in the ground, pedestrians have a terminal class 3GPP class 3 UE.

The following is a list of its main features:
\begin{itemize}
    \item Orbit: LEO at 600~km
    \item Elevation angle: 30º
    \item Frequency reuse: 3
    \item S-band: 2~GHz both uplink and downlink
    \item Bandwidth: 10 MHz downlink and 360~KHz uplink
    \item SINR: 5.5~dB in downlink and 2.5~dB uplink
    \item SE: 1.35 and 1~bps/Hz in downlink and uplink respectively
    \item Bitrate: 13.5~Mb/s and 360~Kb/s DL and UL respectively.
\end{itemize}

\subsection{Intelsat HAPs}

This project is intended to provide connectivity to deep rural areas in Nigeria using HAPs at nominal altitude of 20$\pm$2~km (50~ms approx) covering a fixed area on Earth of 50~km radius (that is 7855~km$^2$). The number of antenna beams on board is 16, hence each beam covers a cell of radius $R_{cell}$ as follows:
$$A_{cell} = \frac{A_{HAP}}{16} \quad \Rightarrow \quad \pi R_{cell}^2 = \frac{\pi R_{HAP}^2}{16}$$
Thus:
$$R_{cell} = \sqrt{\frac{R_{HAP}^2}{16}} = 12.5~km$$

Regarding users on ground, these are considered to follow 3GPP Class 3 UE (that is, 0~dBi antenna gain and 9 dB NF in reception and 23~dBm of EIRP in transmission).

Other important parameters of this NTN include: 
\begin{itemize}
    \item Orbit: HAPs (18-22~km altitude)
    \item S-band: 1.8~GHz
    \item Bandwidth: 13 MHz both DL and UL
    \item SINR: about 21~dB in the center of cell and between 2 and 7.9~dB at the border. 
    \item Link margin: 4~dB for rain fade.
    \item SE: 5.555 bps/Hz for DL and between 0.8 and 1.4~bps/Hz in UL.
    \item Bitrate: ~72~Mb/s for DL and between 10 and 18~Mb/s in UL.  
\end{itemize}

\subsection{Inmarsat GEO IoT}

This GEO project was conceived for IoT applications (NB-IoT standard) in Algeria, where latency is not critical and the IoT applications do not have important bandwidth requirements. The following list shows some of its main features:
\begin{itemize}
    \item Orbit: GEO (38,000~km altitude)
    \item L-band: 1.5~GHz
    \item Bandwidth: 200~KHz for DL and 15~KHz for UL
    \item SINR: not provided
    \item SE: 0.67~bps/Hz for both DL and UL respectively.
    \item Bitrate: 112~Kb/s to 9.33~Kb/s in DL and UL respectively.
\end{itemize}

\subsection{Echostar GEO}

In this project, GEO satellites offer land-mobile and broadband connectivity to rural areas in both Africa and America. Two user equipments are possible: 3GPP Class 3 UE and VSATs for better SINR, allowing higher bitrates. Some of its main features include:
\begin{itemize}
    \item Orbit: GEO (38,000~km altitude)
    \item S-band: 2~GHz
    \item Frequency reuse: 3
    \item Bandwidth: Not specified.
    \item SINR: 15.4~dB and 11~dB in DL and UL respectively for VSATs, and 3 and 0.7~dB for Class 3 UE.
    \item SE: 4 and 2.5~bps/Hz for both DL and UL respectively for VSATs, and 1.2 and 1.0~bps/Hz for Class 3 UE.
    \item Bitrate: Not specified.
\end{itemize}

\subsection{OneWeb LEO}
In this project, LEO satellites offer both ubiquitous connectivity and high-capacity worldwide, allowing seamless integration with terrestrial networks. Flat pannel antennas are considered on the ground for better G/T (between 7 and 9 dB)

\begin{itemize}
    \item Orbit: LEO 1,200 km
    \item S-band: Ku (11.7~GHz DL and 14.5~GHz UL)
    \item Atmospheric loss: 2 dB margin
    \item Bandwidth: Not specified.
    \item SINR: Not specified.
    \item SE: Not specified
    \item Bitrate: 830~Mb/s symmetrical in best case, 140~Mb/s worst case.
\end{itemize}


\subsection{Intelsat GEO HTS}

In this project, a GEO High-Throuhput Satellite (HTS) operating in the Ku band is considered for its use to provide broadband to maritime scenarios in the Mediterranean Sea. VSAT antennas with high G/T values on ground are considered.

\begin{itemize}
    \item Orbit: GEO (38,000~km altitude)
    \item Ku-band
    \item Number of beams: hundreds to thousands
    \item Bandwidth: Not specified.
    \item SE: 0.6~bps/Hz in DL and 1.6~bps/Hz in UL worst case (beam edge); 1 and 1.6~bps/Hz respectively as best case.
    \item Bitrate: Not specified.
\end{itemize}


\subsection{Avanti GEO HTS}

In this project, a GEO HTS is conceived to provide connectivity for connected cars in Western Europe or North America. The car is considered to have on board an antenna with large G/T of 7.4~dB/K.

\begin{itemize}
    \item Orbit: GEO (38,000~km altitude)
    \item Ka-band
    \item Bandwidth: Not specified.
    \item SINR: 2.3 and 4.4 in DL and UL respectively, under clear sky considerations.
    \item SE: 0.9 and 1.3~bps/Hz as best case.
    \item Bitrate: Not specified.
\end{itemize}


\subsection{Hispasat Amazonas 3 GEO}

In this project, several satellites (Amazonas 1, 2, 3 and subsequent) have been launched on GEO orbit to provide connectivity both in rural areas and maritime applications. Some of its features are:

\begin{itemize}
    \item Orbit: GEO (35,786~km altitude)
    \item Ka, Ku and C band
    \item Number of beams: 63 transponders (9 for user and 4 for gateway in Ka band, 33 in Ku band and 19 in C band).
    \item Bandwidth: 54 MHz (Ku and C) and 36 MHz (Ku and C).
    \item SINR: Not specified.
    \item SE: Not specified 
    \item Bitrate: Between 30 and 60 Mb/s.
\end{itemize}

\subsection{Summary table}

Table~\ref{tab:num_examples} shows a summary of the main features of the previous on-going NTN projects.

\begin{table*}[htbp]
    \centering
    \caption{Use cases and demonstration scenarios of~\cite{ngmn_ntn}}
    \begin{tabular}{|c c c c c c |}
    \hline
    Contributor & Orbit & freq & BW & SINR (DL/UL) & SE (DL/UL) \\
    \hline
    Thales & LEO & S (2~GHz) & 10/0.36~MHz & 5.5/2.5~dB & 1.35/1~bps/Hz \\ 
    Intelsat & HAPS (18-22~km) & S (1.8~GHz) & 13/13~MHz & 13~dB & 2.2/2.4~bps/Hz \\ 
    Inmarsat & GEO IoT (38,000~km) & L (1.5~GHz) & 200/200~KHz & NA & 0.6-1.33~bps/Hz \\ 
    EchoStar & GEO 38,000~km & S (2~GHz) & NA & 15(3)/11(0.7)*~dB  & 4/1.2~bps/Hz \\ 
    OneWeb & LEO 1,200~km & Ku & NA & 140 to 830~Mb/s & 4/1.2~bps/Hz \\ 
    Intelsat & GEO HTS & Ku & NA & 0.5-1.9 9 dB~dB  & 0.6-1.9~bps/Hz \\ 
    Avanti & GEO HTS & Ka & NA & 2.3/4.4~dB  & 0.9/1.3~bps/Hz   \\
    \hline
    \end{tabular}
    \label{tab:num_examples}
\end{table*}















\section{Discussion and future work}
\label{sec:conclusions}

This article has briefly overviewed the use of satellites and HAPs for providing connectivity in those areas where fiber cannot reach, namely deep rural areas, mountains, desert and seaside. The foundations regarding link budget analysis and achievable bitrates are also reviewed, along with an introduction to antenna array systems and cellular designs. Finally, some of the most popular emerging satellite constellations and HAP projects are briefly reviewed, along with their characteristics and limitations.











\section*{Funding} 

The authors would like to acknowledge the support of project 6G-INTEGRATION-3 (grant no. TSI-063000-2021-127), funded by UNICO-5G program (under the Next Generation EU umbrella funds), Ministerio de Asuntos Económicos y Transición Digital of Spain.


\bibliographystyle{plain} 
\bibliography{satellites,PONs,sample}

\end{document}